# Structure Analysis using Time-of-Flight Momentum Microscopy With Hard X-rays: Status and Prospects

Olena Fedchenko[1], Aimo Winkelmann[2], and Gerd Schönhense[1]*

*1 Institut für Physik, JGU Mainz, D-55099 Mainz, Germany*

*2 Academic Centre for Materials and Nanotechnology (ACMiN),*

*AGH University of Science and Technology, 30-059 Kraków, Poland*

**Abstract**

X-ray photoelectron diffraction (XPD) has developed into a powerful technique for the structural analysis of solids. Extension of the technique into the hard-X-ray range (hXPD) gives access to true bulk information. Here we give a status report on hXPD experiments using a novel full-field imaging technique: Time-of-flight momentum microscopy (ToF-MM). A special variant of ToF-MM is capable of recording high kinetic energies (up to >7keV) and enlarged k-fields-of-view. We present applications that are specific for high kinetic energies. The strong site specificity of hXPD is exemplified for $NbSe_2$, the cubic-to-tetragonal transition in $SrTiO_3$ and the zinc-blende structure in epitaxial GaAs films. Bloch-wave calculations show a very good agreement with experiment and reveal fingerprint-like signatures of emitter sites in host lattices. We show a dopant-site analysis in two semiconductors (Mn in GaAs and Te in Si). Hard-X-ray ARPES plus core-level hXPD enable eliminating the strong diffraction signature imprinted in HARPES maps.

**Keywords:** Photoelectron diffraction, XPS, Hard X-rays, Bloch-wave method

## 1. Introduction

The discovery of X-ray photoelectron diffraction (XPD) dates back to the pioneering work of Siegbahn et al.[1] and Fadley and coworkers.[2,3] Today, XPD is an established method for an effective retrieval of structural information.[4,5,6,7,8] Strongholds of XPD are its capabilities to analyze adsorbate geometries [9] and its close relation to photoelectron holography.[10,11,12,13] Even a spin-polarized version has been developed by the Fadley group.[14,15] Element selectivity, even sensitivity on the chemical state of an atom, and high site specificity can be exploited as unique fingerprints of atomic sites in compounds[16]. By combining angle-resolved photoelectron spectroscopy (ARPES) and XPD in a single experiment, electronic and structural information can be obtained quasi-simultaneously and at identical conditions (kinetic energy of the photoelectrons, size and position of the probing spot and probing depth).[17,18]

The last decades have seen dramatic advances in the technical performance of photoemission experiments, both on the source and detector side. Hemispherical analyzers (HSAs) are the workhorses of photoelectron spectroscopy. After almost a century of development, HSAs have reached excellent performance with high energy resolution (down to < 1 meV)[19, 20, 21, 22] and high angular resolution (equivalent to a momentum resolution of 0.003 Å$^{-1}$).[23] In the classical mode of operation [($E_{kin}$,θ)-mode] the exit plane of a HSA hosts a 2D detector, which can record an energy interval of typically 10% of the pass energy and an angular interval of +/-7° in the high-angular-resolution mode. Larger angular ranges can be recorded using wide-angle lenses or sequentially via sample rotation. For XPD experiments, the full 2π solid-angle interval can be captured by sequential measurements scanning polar and azimuthal angles.[4,17,24]

A much younger alternative way of high-performance photoemission experiments is the momentum microscope (MM). These instruments exploit a basic principle of optics, namely that the backfocal plane of an objective lens hosts a reciprocal image (in particle optics this corresponds to nothing else than a momentum image). The MM family comprises several types of instruments, the energy-dispersive type using either a tandem arrangement of two hemispherical analyzers[25, 26, 27] or a single hemispherical analyzer[28, 29]. The second way of energy discrimination employs time-of-flight detection, leading to the ToF-MM[30]. A third approach is a hybrid instrument of single hemisphere plus additional ToF section [31, 28]. This development is based on the time-resolving imaging technique originally developed for the ToF-PEEM[32, 33]. The key feature of all MMs is their large solid-angle acceptance, exploiting the full-field-imaging properties of a cathode lens. Capturing energy intervals of several eV simultaneously, ToF-MMs record large sections of ($E_{kin}$,**k**) parameter space without sample rotation or angular scanning.

Naturally, time-of-flight energy recording needs pulsed photon sources. Thanks to their perfectly-periodic time structure, synchrotron radiation, pulsed lasers, higher-harmonic-generation (HHG) sources and fs-pulsed sources like free-electron lasers (FELs)[34, 35] or the upcoming attosecond facilities[36, 37] fit to the ($E_{kin}$,$k_x$,$k_y$) recording scheme of a ToF-MM.

An alternative technique of accessing large solid-angle intervals employs a combination of grids. Eastman et al. developed the first instrument of this kind with an ellipsoidal electron mirror as low pass and spherical grid as high pass filter stage as early as 1980.[38] Soon thereafter, Daimon published an improved design[39], which has led to the *display-type spherical mirror analyzer* DIANA.[40] This instrument launched a very successful development in Japan, see recent reviews[10, 11] and references therein. Being both full-field imaging instruments, display-type analyzers and ToF-MM bear much resemblance. The main difference is the observed angular range, covering the full $2\pi$ solid angle for DIANA, and a momentum range of diameter ~20 Å$^{-1}$ for the ToF-MM, corresponding to a polar angular range of up to 0-15°, depending on kinetic energy. The relative angular resolution is comparable: 0.5° for DIANA[41] and 0.034° for the ToF-MM[16]. Principal limitations are posed by the lens aberrations, limiting the maximum accessible angular range for the ToF-MM and the micro-lens action of the grid meshes[42] for display-type analyzers.

This progress concerning XPD experiments comes along with the availability of soft- and hard-X-ray synchrotron beamlines and forthcoming X-ray free electron lasers. At the new beamline P22 of PETRA III a total energy resolution (combined resolution of ToF analyzer and photon bandwidth) of 62 meV at 6 keV (resolving power ~$10^5$) has been reached.[43, 44] At the free electron laser FLASH (DESY, Hamburg) a similar ToF momentum microscope is operated[45] and the first time-resolved XPS and XPD experiments with fs resolution at soft-X-ray energies have been performed.[46, 47, 48, 49]

Here we give a status report on the application of the ToF-MM method for full-field imaging of XPD patterns in the hard-X-ray range (hXPD) for geometric structure analysis. These experiments employ a special variant of ToF-MM, capable of recording high kinetic energies (up to >7 keV) and enlarged k-fields-of-view. In particular, we discuss those properties of hXPD, which are specific for high kinetic energies.

One principal advantage of using photoelectron diffraction for structural analysis is that – unlike in conventional single-scattering X-ray diffraction - XPD circumvents the preconditions of Friedel's rule by pronounced multiple scattering. As an application, hXPD proved that the zinc-blende structure (with a missing inversion center) persists with macroscopic order in thin epitaxial GaAs films[17]. Another special property is the strong site specificity of hXPD, demonstrated for the transition-metal dichalcogenide NbSe$_2$ and strontium titanate. Calculations using the Bloch-wave approach to hXPD show a stunning agreement with experiment and reveal fingerprint-like signatures of different emitter sites in a host material. As examples, we show a dopant-site analysis in two semiconductors (Mn in GaAs and Te in Si). Another application is the observation of structural phase transitions as exemplified for the cubic-to-tetragonal transition in SrTiO$_3$. Off-normal emission at large polar angles can be captured using the zero-extractor-field mode of the front lens. Combining hard-X-ray ARPES (HARPES) with core-level hXPD, it is possible to eliminate the strong diffraction signature imprinted in bulk band maps taken with hard X-rays. Identical settings (kinetic energy, *k*-scale) enable a pixel-by-pixel evaluation / correction procedure (difference or ratio images, removal of imprinted hXPD modulations) of the patterns. Parallel ($k_x,k_y$)-imaging and ToF energy recording effectively counteract the rapidly dropping cross sections[50] and increasing phonon scattering with increasing photon energy in the hard X-ray range.

## 2. Full-field imaging of photoelectron diffractograms
*2.1 Photoelectron momentum microscope: an effective tool for hXPD and HARPES*

A ToF-MM detects the photo-emitted electron intensity $I$ ($E_B,k_x,k_y$) as a function of parallel momentum $k_x$ and $k_y$ and binding energy $E_B$.[30] Fig. 1(a) illustrates the functioning principle. The first imaging element (objective lens) of the zoom optics behind the sample is a so-called cathode lens, of which the sample itself is an integral element. This lens forms an achromatic momentum image in its backfocal plane (BFP). This image is magnified to the desired size by a two-stage zoom optic comprising several lens groups. The photoelectron distribution is dispersed in time-of-flight by a field-free low energy ToF drift section and recorded in an (x,y,t)-resolving delay-line detector (DLD).[51] A movable and size-selectable field aperture in the first Gaussian image plane allows a precise definition of the analyzed region-of-interest (ROI) down to the sub-micrometer range (see Fig. 6 in [31]). By switching lenses to the real-space imaging mode (PEEM mode), this aperture and its position on the sample surface can be viewed on the detector. The sample is mounted on a He-cooled hexapod manipulator for precise six-coordinates adjustment (*x,y,z,*$\theta_x,\theta_y,\phi$).

The efficiency of the three-dimensional recording architecture of ToF-MMs is defined by the time-resolving image detector. Delay-line detectors[52, 53] or the upcoming solid-state pixel-array detectors[54, 55] are characterized by time resolutions in the 50-200 ps range. Hence, for ToF-based experiments not only the pulse length but also the pulse period of the photon source is crucial because the ratio between duration and period determines, how many time slices can be resolved in the time gap between adjacent photon pulses. We show experiments using the 40-bunch mode of the storage ring PETRA-III (at DESY, Hamburg) with 192 ns period. Given the time resolution of ~180ps, the resulting number of resolvable time slices is ~1000, being sufficient for high-resolution spectra. Beamline P22[43] provides hard X-rays in the range from 2.5 to >10 keV in a photon spot of down to <20 μm and at a resolution down to 60 meV for the Si(333) monochromator cystal.[44]

In a first model study, hard x-ray photoelectron diffraction (hXPD)-patterns have been recorded for a graphite single crystal at many photon energies.[56] The measurement yielded detailed Kikuchi-type diffraction patterns, which showed an excellent one-to-one agreement with simulations using the Bloch-wave approach[57], see Fig. 1(b-h). The excellent agreement (including the faint fine structure in Kikuchi bands and zones) of calculated and measured hXPD patterns suggests to use this method as sensitive tool for structural analysis. Full-field imaging is highly efficient; in the case of the strong C 1s core-level signal of graphite a hXPD pattern is recorded within 10 min. The Bloch-wave simulation method is highly efficient as well, because it allows to calculate the Kikuchi patterns in terms of Bragg scattering of electrons from point sources inside a crystal using a rather small number of Fourier components (<100) of the scattering potential (equivalent to sets of lattice planes), as described in Ref. 57. This approach emphasizes the long-range order and is thus complementary to cluster codes[58, 59, 60], which have their strength in modeling the nearest-neighbor configuration.

Hard-X-ray XPD patterns were taken in the geometry sketched in Fig. 1(a) for 20 different photon energies in the range of 2.840–7.283 keV. The signal from the C 1s core level was selected by setting the proper sample bias such that the point of best focussing coincides with the maximum of the C 1s peak. Graphite turned out to be favourable as the first example of Kikuchi-band observation using the new method. C 1s is the only core level; hence its signal appears prominent above a very low background. Moreover, due to its low Z the scattering factor is low which leads to pronounced Kikuchi patterns with straight lines and bands clearly visible in real time (exposure 1 s). The *k*-space metric of Kikuchi bands crossing the centre of the XPD pattern of a low-index crystal surface is simple and allows reconstructing the size of the projected Brillouin zone. As described in Fig. 1 (h) the central bands originate from sets of lattice planes oriented perpendicular to the surface. In the case of graphite the dominating six-fold-symmetric Kikuchi pattern originates from reflection at <110> lattice planes. The width of the prominent (main) Kikuchi band is given by the corresponding reciprocal lattice vector $2|G_{110}| = 4\pi/1.228$ Å = 5.166 Å$^{-1}$.

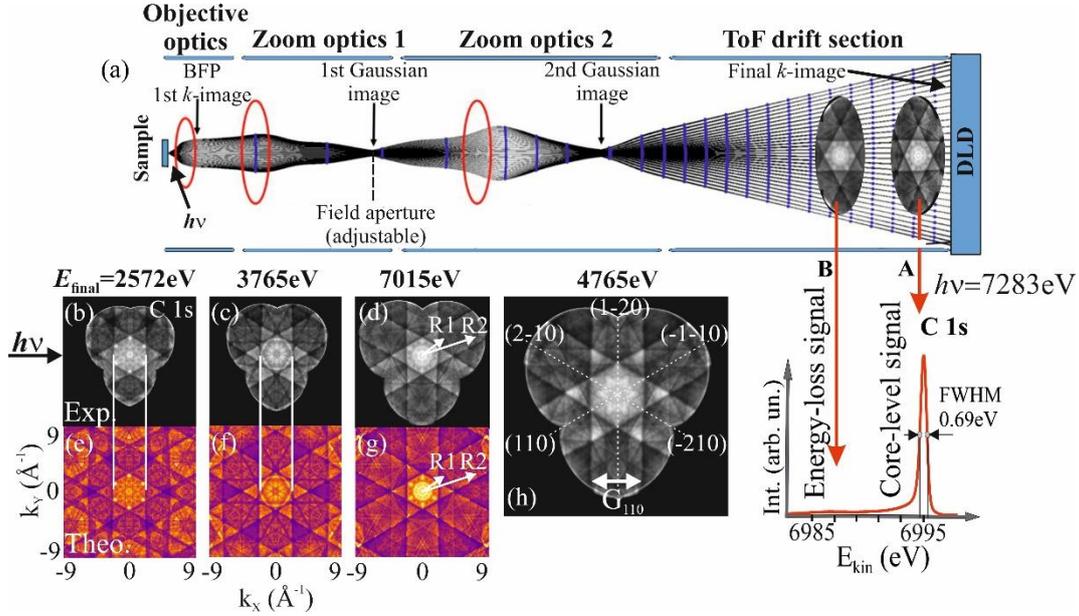

**Fig. 1.** (a) Schematic cross-section of the ToF MM with calculated trajectories. Comparison of measured (b-d) and calculated (e-g) C 1s hXPD-patterns of graphite; calculations based on the Bloch-wave approach. (h) Measured Kikuchi pattern of Graphite near the [001] zone axis for $E_{final}$ = 4.765 keV [(b-h) from Ref. 56]. Photon impact is in the horizontal plane at 68° off-normal; the projection of the incident ray is indicated by an arrow in (b).

Sharp bright and dark lines shift with energy; regions of Kikuchi band crossings near zone axis exhibit a filigree structure, which varies rapidly with energy. The fine structure in the Kikuchi pattern, including the dark features inside the central zone originate from reflection at lattice planes with small spacing and correspondingly large Bragg angles. We also note that circular features [R1, R2 in Figs. 1(d,g)] and other complex intensity distributions are formed as envelopes of lines.

Parallel energy recording via ToF yields energy bands of several eV width in parallel in a single acquisition. Hence, it is possible to extract several different diffractograms from a single 3D data array, as demonstrated in Fig. 1(a) and the corresponding ToF spectrum. Integrating over a small energy interval close to the maximum of the C 1s signal (here at a kinetic energy of 6995 eV) yields the core-level diffractogram (A).

Integrating over an energy interval in the low energy tail (here around a loss-energy of 8 eV) yields the 'energy-loss hXPD diffractogram' of those electrons that have been inelastically scattered (B). Here $E_{kin}$ is the kinetic energy in vacuum, relevant for photoemission spectra, whereas the final-state energy $E_{final}$ is the kinetic energy of the electrons inside of the material (details in Sec. 2.2).

Figs. 1(b-g) reveal that, without exception, all details contained in the measured diffractograms are found in the calculation as well. The only significant differences are that the experimental patterns are slightly blurred and some details appear at different relative brightness. In order to observe the [001] zone axis [central hexagon, i.e. the intersection region of all three (110) Kikuchi bands] fine structure in Figs. 1(b,c,d) with high resolution, we have chosen a $k$-field-of view of 14 Å$^{-1}$, about twice the width of the main Kikuchi band. Translated to real-space (angular) coordinates this corresponds to a polar angular range of only 0°–9°, being much less than typical angular ranges in previous XPD experiments. However, our $k$-resolution of 0.025 Å$^{-1}$ is equivalent to an angular resolution of 0.034°, which is 1–2 orders of magnitude better than reported in previous work (all values for 7 keV kinetic energy).[56]

The experimental patterns were obtained in one measurement with eccentric adjustment of the field of view, in order to reach a larger $k_\parallel$. The off-center initial pattern was mirror-flipped with respect to the horizontal and vertical to show a larger total field of view for comparison with theory. No special processing for the patterns was used, except for a rotation by small angles before the first flipping to avoid artefacts along the overlapping line. No further corrections have been applied.

The front lens of the high-energy ToF-MM enables a *zero-extractor-field mode*. The important action of the extractor in classical PEEM and LEEM is to form a virtual image of the sample with high kinetic energy and small angular divergence. In hard-X-ray MM the starting energy is already very high and the recorded angular range is small. Moreover, the small photon spot of ~20 μm ensures a large depth of focus in the k-image. The zero-field mode is particularly useful for observing large polar angles θ, see example for θ=60° [Figs. 5(j-o)]. Further important applications of this zero-field mode are experiments with strongly 3D-structured samples or sample mounts with protruding contacts or actuator elements on top.

*2.2 What do we observe in soft- and hard-X-ray photoelectron diffraction?*

Before discussing the examples, it is worthwhile to consider the differences of XPD in the soft- and hard-X-ray range. Here we briefly recall the basics of X-ray photoemission, emphasizing the intrinsically different nature of the final states at low and high kinetic energies. As extrinsic effect, phonon-mediated quasi-elastic scattering processes become significant at high energies and finally their intensity even dominates over the direct photoelectron wave. This is especially important for valence-band mapping (HARPES), as will be discussed in Sec. 4.

Core-level XPS describes transitions from a localized initial state to an outgoing spherical-wave final state. This electron wave is diffracted at the lattice, which is the concept of Pendry's model of the 'multiple-scattering' photoemission final state.[61, 62, 63] Since it was first discussed for low-energy photoemission, it is also termed 'time-reversed LEED-state' (LEED: low-energy electron diffraction). In an *E-vs-k* scheme, X-ray photoemission is described by direct transitions into quasi-free-electron-like final states with parabolic dispersion of the final-state energy $E_{final}$ (inside of the material) versus final-state momentum $k_f$ [$E_{final}$ in eV, $k_f$ in Å$^{-1}$]:

$$|k_f| = 0.512\ [(m_{eff}/m_e)\ E_{final}]^{1/2} \quad \text{with} \quad E_{final} = h\nu - E_B + V_0^* \quad (1).$$

The inner potential $V_0^*$ is referenced to $E_F$. At X-ray energies, the effective mass $m_{eff}$ differs only slightly from the free-electron mass $m_e$. For tungsten $m_{eff}/m_e$=1.07 at $E_{final}$= 1 keV[30] and 1 at 6 keV[64]. In principle, $m_{eff}$ could be (weakly) direction-dependent. In particular, very close to the BZ boundary (where Bragg reflection occurs) the deviation from parabolic dispersion should be visible. However, to the best of our knowledge no experimental evidence exists for varying $m_{eff}(k)$ in the X-ray range. The behaviour of $m_{eff}$ is related to the concept of 'dispersion surface' in TEM, for details, see Ref. 65. The effect of the inner potential is visible in low-energy electron diffraction but largely negligible at hard X-ray energies [66]. Valence-band photoemission is more sensitive to variations of $V_0^*$ and $m_{eff}$. In principle, both quantities can be determined from a measurement sequence at different photon energies [30].

The *direct photoemission intensity* $I_{direct}$ is proportional to the transition probability *P* from an initial state $\varphi_i$ to a final state $\varphi_f$ as described by *Fermi's Golden Rule*. The operator $\Delta$ represents the electromagnetic field and the $\delta$–function accounts for energy conservation. The polarization of the photon beam enters the operator and gives rise to the *symmetry selection rules*.[67]

$$I_{\text{direct}} \propto \quad P = \frac{2\pi}{\hbar} |<\varphi_f|\Delta|\ \varphi_i>|^2 \delta(E_f - E_i - h\nu) \qquad (2).$$

The photoelectron wave 'feels' the lattice surrounding the emitter atom. Here we must distinguish two different regimes: In the low-energy regime at few 100 eV, the photoelectron wavelength is of the order of the nearest-neighbour distances and the IMFP is of the order of 1 nm. The wave field is governed by multiple-scattering events at individual atoms in a cluster surrounding the emitter atom. In this regime the immediate vicinity of the emitter is probed, the observed intensity modulation arises from the interference of the outgoing direct wave with the scattered waves. Hence, the kinetic-energy range of several 100 eV can be termed the *holographic regime*.

Matsushita et al. proved that the hologram of diamond at $E_{kin}$= 900 eV can be described with very good agreement by a calculation assuming a cluster of 1 nm radius (see Fig. 2 of Ref. 41). The authors discuss the differences compared to classical Kikuchi patterns and introduce the terminology 'quasi-Kikuchi-band', in order to distinguish it from the high-energy scenario. Their program code allows varying all relevant parameters; best fit is obtained for IMFP = 1.2 nm and a thermal-vibration amplitude of 3.4 pm (at room temperature).

In the high-energy regime, we face a different situation. Let us analyze the example shown in Figs. 1(b)–(d) quantitatively. In the sequence from 2.57 to 7.01 keV [full sequence, see Ref. 56], two trends are important: the *photoelectron wavelength* drops from 25 pm to 14.6 pm and the *IMFP* increases from ~ 4 to ~10 nm.[68)] Assuming that the effective probe volume is defined by roughly twice the IMFP, its radius increases from 8 to 20 nm in this energy range. Inserting the density of graphite (2.25 g/cm$^3$), we arrive at a number of about 3x10$^5$ to 4x10$^6$ carbon atoms in the active probe volume, for $E_{kin}$= 2.57 to 7.01 keV, respectively. This is the *regime of Bragg reflection of Bloch waves at sets of lattice planes*. Being sensitive to the local atomic configuration (even for non-periodic systems) and the position of an emitter atom with respect to an extended periodic lattice, the methods exploring the *holographic regime* and the *Bragg-reflection regime* are nicely complementary.

Owing to the small photoelectron wavelength, the Bragg angles for the low-index (i.e. widely-spaced) lattice planes are very small. At $E_{\text{final}}$ = 7 keV we have $|k_{\text{final}}|$ = 42.9 Å$^{-1}$ and for the <1 1 0>-type of lattice planes of graphite (Fig. 1(h)) we derive a Bragg angle $\theta_B$ of only 3.4°. The Bragg-reflected electrons appear in the momentum image at a distance of 2.56 Å$^{-1}$ from the centre of the pattern. XPD diffractograms at such high energies show a combination of *k*-space features (i.e. Bragg reflection operative at Kikuchi band edges, with the width of the bands given by multiples of reciprocal-space vectors), as well as long-range real-space lattice features such as the center lines of the Kikuchi bands as projections of lattice planes (hkl) and the crossing points of Kikuchi band centers as projections of crystallographic directions [uvw]. The hXPD scenario bears many similarities to Kikuchi diffraction, being a powerful tool for structural analysis in scanning and transmission electron microscopy (SEM and TEM, respectively)[69, 70, 71]. The basics of hXPD calculations will be described in Sec. 2.4.

Here the long-range order of the lattice is crucial. Unlike the low-energy situation, there is no significant amplitude of the 'direct wave' at the detector, and the Kikuchi bands are governed by multiple scattering in the long-range lattice geometry, which makes the local holography concept a less efficient description. However, due to the small photoelectron wavelength, the diffraction dynamics depends sensitively on the local perspective of the emitter atom on the lattice. In compounds, this 'perspective view on the lattice' is generally different for the various constituents and, in turn, the hXPD patterns of different atom species in compounds are different. Instead of gaining local information from a hologram, we retrieve the position of the emitter atom from its fingerprint-like hXPD diffractogram, i.e. from the specific contrast features in the multiple-scattering Kikuchi bands.

This site-specificity is explained in the schematic picture in Fig. 2 (a,b) and example (e-g). The outgoing photoelectron waves are centered at two emitter atoms A and B, located on different, non-equivalent lattice planes (full and dashed). The example shows diffractograms for the transition-metal dichalcogenide 2H-NbSe$_2$. The Nb 4p$_{3/2}$ (e) and Se 3p$_{3/2}$ (f) core-level hXPD patterns show the same width of the Kikuchi band [2G$_{110}$ = 3.65 Å$^{-1}$], which is defined by the lattice structure. However, the fine structure in (e,f) is markedly different, as quantified in the ratio image Fig. 2(g). In order to exclude dynamical effects, these images were recorded at identical final-state energies of 3.617 keV by choosing the appropriate photon energies. The structure model Fig. 2(h) illustrates the different positions of Nb and Se in the lattice of the compound. In comparison with the situation in TEM and SEM, core-level hXPD can be considered as a *more controlled Kikuchi-process* starting from an internally created (photo-) electron point source, without the various possible parasitic effects caused by inelastic scattering of an incident electron beam.

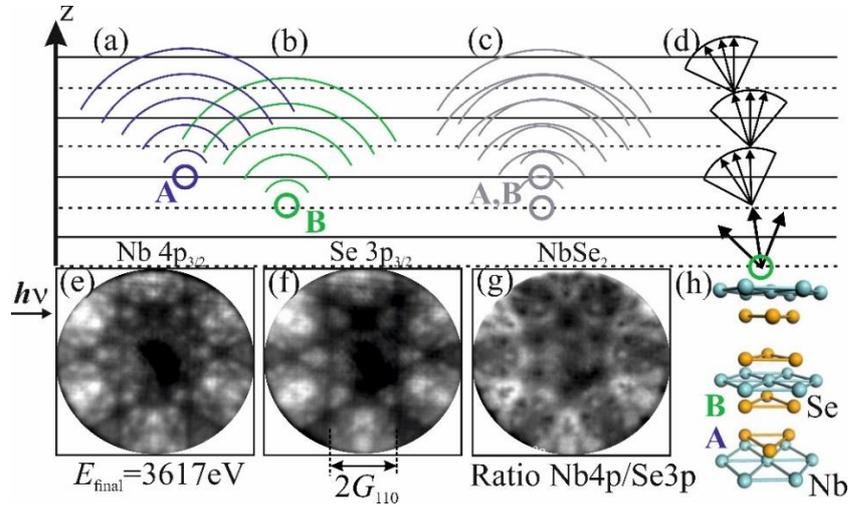

**Fig. 2.** Schematic of different contributions to photoelectron diffraction: a) the photoelectron wave (the 'Pendry multiple-scattering final state') showing the fingerprint-like hXPD pattern corresponding to emitter atom A; b) same for a photoemitter atom B on a non-equivalent site. These two channels dot not interfere and can be separated by the core-level binding energies of A and B. (c) Quasi-elastic thermal diffuse and other inelastic scattering kicks electrons out of the final-state wave field related to a specific emission site. The scattered electrons constitute a new wave field centered at the scatterer (here A or B), with characteristic weighted average Kikuchi pattern. (d) Cascade-like sequences of scattering events may occur at high kinetic energies. (e-g) Experimental example: The core-level hXPD patterns of Nb $4p_{3/2}$ (e) and Se $3p_{3/2}$ (f) in 2H-NbSe$_2$ recorded at identical final-state energy of 3.617 keV are markedly different, best visible in the ratio image (g). This reflects the different local geometry of the two atom species, see structure model (h).

*2.3 The role of quasi-elastic scattering*

Inelastic scattering events can destroy the coherence with the Pendry final-state wave, leading to a corresponding localization of the scattered electron. The electron 'drops out of the wave field', but its energy is almost retained except for a small change of the order of the atomic recoil energy. The classical point-mass elastic scattering picture of a recoiling free atom due to electron impact corresponds quantum-mechanically to a coherent multi-phonon excitation in a lattice[72, 73, 74]. Immediately after the scattering event, the recoil energy is stored in a coherent phonon superposition that resembles the initial local displacement of the recoiling scattering atom. Due to the elastic forces, which pull the displaced atom back towards its equilibrium position, the local lattice displacement then evolves with time into a delocalized thermal excitation characterized by the vibrations of the neighboring atoms. In this way, the recoil energy ends up as thermal energy of the lattice. Such quasi-elastic events gain importance with increasing kinetic energy and increasing temperature. We adopt the terminology *thermal diffuse scattering* from electron microscopy, where such effects have been studied in detail (e.g. Ref. 75 and references therein). Since the quasi-elastically scattered electron wave can subsequently also be coherently diffracted at the lattice again, its momentum pattern is by no means completely 'diffuse', but shows the diffraction fingerprint of the site of the scatterer. The electrons localizing out of the final-state wave field by thermal scattering constitute a new wave field, starting at the scattering sites as sketched in Fig. 2(c). In case of several non-equivalent sites, the total pattern is the (incoherent) sum of several site-specific diffractograms, weighted by the scattering cross sections.

With increasing kinetic energy, the mean free path increases[68] and cascade-like sequences of scattering events can occur[75] as schematically sketched in Fig. 2(d). This behaviour is identical to the classical Kikuchi process in electron microscopy. Instead of the primary electron beam in SEM or TEM, here the initial state is the photoelectron wave. In addition to the Kikuchi scenario, contributions from *Laue diffraction* have been observed in valence-band photoemission (cf. Sec. 4). Strong local intensity enhancements, confined to small sections of (*E*,***k***) parameter space, correspond to umklapp processes on the final-state energy isosphere and can be visualized in terms of a graphical intuitive model.[76]

Assuming that no electron gets lost (we consider only elastic or quasi-elastic processes), the relative strengths of the direct and non-direct (i.e. phonon-mediated) intensity in angular- or *k*-resolved HAXPES are governed by the simple law:

$$I_{total} = W(T)\, I_{direct} + [1 - W(T)]\, I_{non-direct} \qquad (3),$$

with the Debye-Waller Factor *W*(T) depending on the reciprocal lattice vector *G* (momentum transfer) and the mean (thermal) atom displacement *u* according to:

$$W(T) = exp\left(-\frac{1}{3}\vec{G}^2 \langle u^2(T) \rangle\right), \qquad \langle u^2(T) \rangle \propto \frac{T}{M_{atom}\, \theta_{Debye}} \qquad (4).$$

The measured momentum patterns of core-level signals in the hard-X-ray range thus comprise several contributions: The photoelectron wave is modulated by the fingerprint-like diffraction pattern corresponding to the emitter atom (Pendry final state), see e.g., constituents of NbSe$_2$ in Fig. 2(e,f). The underlying diffraction process can be considered as half a Kikuchi process with a missing initial backscattering step. In turn, such patterns strongly resemble Kikuchi patterns and the calculation employs the same code as used to calculate Kikuchi diffractograms in SEMs and TEMs[77]. Such fingerprints can be utilized, e.g., to determine sites of dopant atoms in a host lattice (see Sect. 2.4). In addition, an incoherent contribution from quasi-elastically and inelastically scattered electrons is superimposed. This contribution increases with temperature and energy, and leads to excess-deficiency Kikuchi lines in the general case of an anisotropic angular intensity distribution, even for a totally incoherent electron distribution.[78]

The above-mentioned quantitative estimation indicates that the prominent features in Figs. 1(b-d) and 2(e-g) originate from Bragg reflection under shallow impact angles; in other words, from near-forward scattering. A look at the cross sections for elastic scattering (the diffraction channel) and thermal diffuse scattering (the phonon-mediated channel) shines further light into this scenario. A calculation for Si by Wang reveals that these two single-atom cross sections are markedly different (see Fig. 3 in Ref. 75; note the log scale!). When plotted against the momentum transfer, the elastic cross section shows a steep rise with maximum in forward direction. With increasing momentum transfer, it shows a dramatic drop (in case of Si by 5 orders of magnitude). The thermal diffuse cross section is much smaller in forward direction and exhibits a weaker drop with increasing momentum transfer. The phonon-mediated channel thus shows no pronounced directional dependence. The dominant forward scattering characteristic alongside with the large number of scatterers in the probe volume explains the richness in details in patterns like Figs. 1(b-g).

Next to phonon-mediated scattering, there are further inelastic processes like energy-loss scattering due to electronic or plasmonic excitations, or e-e scattering involving an itinerant electron of the valence bands. The latter process, leading to a large energy loss of approx. half the kinetic energy, is responsible for the inelastic secondary-electron cascade. In general, such processes have a different probability distribution in real space. In turn, we observe regions where the electron background shows a Kikuchi signature and other regions where it looks unstructured;[79, 80] an example is shown in Fig. 8(o).

Another loss process is the *recoil effect*, originating from the large photoelectron momentum (e.g. $|k_{final}|$ = 43.8 Å$^{-1}$ at $E_{final}$ = 7.3 keV). In order to fulfil the total momentum balance in the photoemission process, this electron momentum is causing a recoil momentum transferred to the emitter atom. For light atoms, this momentum transfer can be observed as an energy shift to lower kinetic energy in the 10 meV range. Since this shift depends on the atomic mass [81], recoil spectroscopy can yield information on the local environments of the emitter atoms in a material [82]. We did not observe a recoil effect in the studies discussed. The recoil would lead to a change in the XPD pattern, if the effective source position in the unit cell would change by large amounts (e.g. 10% of the unit cell dimensions). If this change were random, the XPD pattern would mainly have reduced contrast. The timescale of the photoemission process is much shorter than the time for the movement of the atoms. One thus can speculate that the influence of any recoil effects on the source position of the photoemission event will be hard to detect. The atom moves much slower than the photoelectron leaves the emission site.

*2.4 Calculations using the Bloch-wave approach to hXPD*

Photoelectron diffraction is a special case of the general problem of electron emission from sources inside a crystal for which a general description can be given using the dynamical theory of electron diffraction[83, 84]. The applicability of many-beam Bloch wave simulations to bulk photoelectron diffraction has been demonstrated in Refs. 57 and 6. For basic geometrical interpretations, the key concept of diffraction patterns from internal sources still remains the Bragg reflection condition for the crystal lattice. For a given lattice plane, the Bragg reflection condition with Bragg angle $\theta_B$ is fulfilled for incident and reflected wave vectors on cones with half opening angle of 90°-$\theta_B$, with the lattice plane normal as the symmetry axis of these so called 'Kossel cones'.[83]

Depending on the specific geometrical projection conditions of the measurement, the Kossel cones are imaged as the corresponding conic sections in a projection plane. When measuring diffraction features in $k_x,k_y$-space, Kikuchi bands with their lattice plane normals (*hkl*) in the projection plane will show a constant width corresponding to the reciprocal lattice vectors $G_{hkl}$. Kikuchi bands of inclined planes will vary according to the respective *z*-component of the $G_{hkl}$ in the projection *k*-space.

The advantage of the Bloch-wave model lies in the effective description of materials with perfect long-range order. The advantage of spherical-wave cluster calculations is their capability of treating non-3D-periodic systems like adsorbate structures. Both methods are thus equally powerful with complementary application scenarios. When fully converged, both approaches should, in principle, yield identical results.[6]

Fig. 3 presents calculated Kikuchi diffractograms for various positions of emitter atoms in the Si unit cell (a). Si was chosen because of its high technical relevance. The Bloch-wave approach enables calculating single emitter atoms (b,f-i), but also the superposition patterns of multimer emitters, here for the example of a linear trimer placed on the space diagonal <111> (c), a triangular trimer (d,j), and a 7-atom cluster (e,k). In (b-d) averaging over all four <111>-axes mimics the statistical distribution in a real system. (e) shows the pattern for an individual 7-atom cluster, sketched in (k). Note the strong differences between monomer (b), trimers (c,d) and cluster (e). Calculations for monomers on different positions are shown for a substitutional site (f), a hexagonal interstitial (g), a tetrahedral interstitial (h), as well as a fully arbitrary position (i). For the particular case of Te dopants in Si, we can compare the calculated results with first experimental data (cf. Sec. 3.3). The results open a new avenue towards a detailed structural analysis with high site-specificity, interesting for many materials whose properties are tailored via doping.

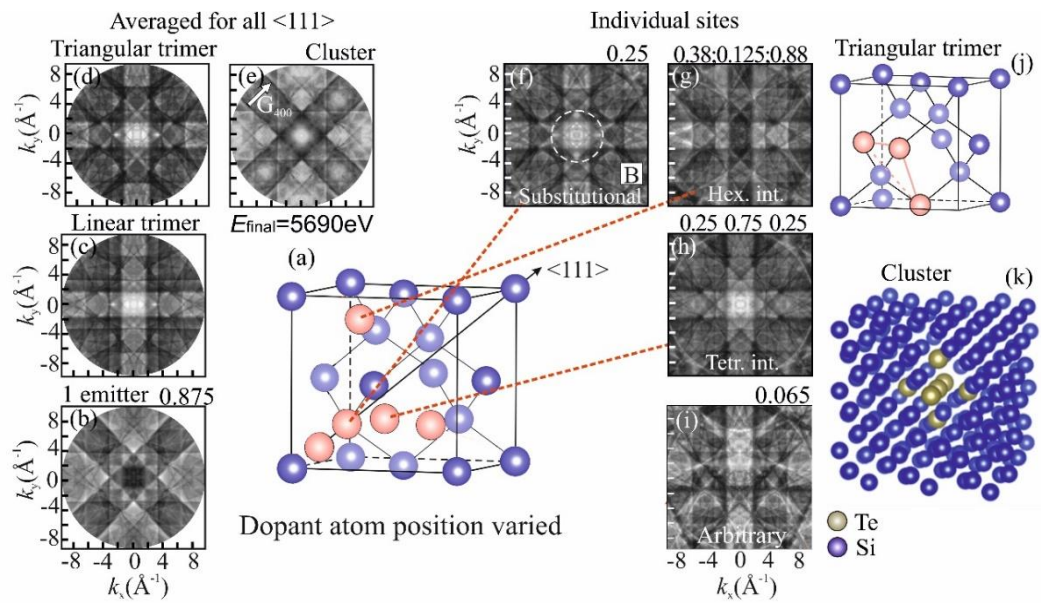

**Fig. 3.** Calculated hXPD fingerprints for different emitter sites in the unit cell of Si (a). (b-d) patterns for linear (b) and triangular trimers (c) and 7-atom clusters (d), all averaged over the four space diagonals <111>. (e) Pattern for an individual 7-atom cluster. (f-i) Monomers on different sites: substitutional (f), hexagonal interstitial (g), tetrahedral interstitial (h) and a fully arbitrary position (i). Three numbers denote the Cartesian coordinates in the unit cell, single numbers refer to the position along the <111> axis. (j,k) sketch of the triangular trimer and the 7-atom cluster. All calculations for $E_{final}$ = 5.69 keV (after Ref. 16).

## 3. Core-level hXPD

*3.1 Lattice analysis: missing inversion symmetry in thin films with zinc-blende structure*

Using hXPD and MM we demonstrate the conservation of broken inversion symmetry and band structure tailoring for high-quality molecular-beam-epitaxy-grown GaAs films. Figure 4 shows examples for measured and calculated hXPD patterns of GaAs [zinc-blende structure (a)] in comparison with Si [diamond structure (f)], uncovering the striking differences of two similar lattices without (b-e) and with inversion symmetry (g). Photon energies of 3–5 keV ensure that the results are not corrupted by surface effects, which are known to be strong in semiconductors. The missing inversion center of the GaAs host lattice leads to fingerprint-like hXPD signatures of As and Ga sites.[17] The measurements were performed with the characteristic parameters: k-field diameter ~10 Å$^{-1}$, *k*-resolution down to 0.025 Å$^{-1}$ (0.034° angular resolution), kinetic energy up to 7 keV, time-of-flight energy resolution ~40 meV (measured at hν = 6 keV photon energy), and photon bandwidth 330 meV (for 3300 eV) for the Si(111) monochromator crystal.

The hXPD patterns in Fig. 4 exhibit a pronounced system of Kikuchi bands and lines with a rich fine structure. The experimental and theoretical diffractograms show a one-to-one agreement of the overall structure, the widths of the observed bands, and the positions of the lines.

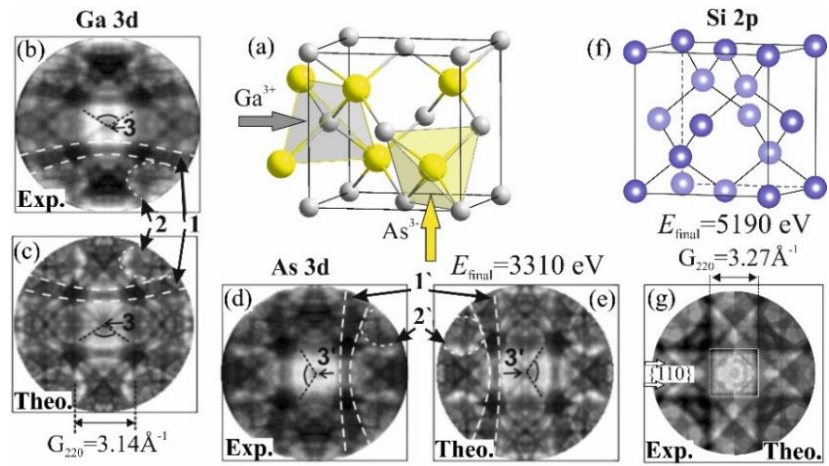

**Fig. 4.** (a-e) hXPD analysis of a GaAs thin film (zinc-blende structure), revealing that the missing inversion symmetry persists in the film. The zinc-blende structure is sketched in (a); note the different orientations of the Ga$^{3+}$ and As$^{3-}$ coordination tetrahedra marked by arrows. Measured and calculated diffractograms for Ga 3*d* and As 3*d* are displayed in (b,c) and (d,e), respectively. (f) Lattice structure and (g) analogous hXPD analysis for a silicon crystal; (g) displays both the measured (left half) and calculated Si 2*p* patterns (right half, V$_0$=15eV). This pattern is fourfold-symmetric, reflecting the inversion symmetry of the Si lattice (f). [(a-e) from Ref. 17; (g) from Ref. 16].

Close inspection of the Ga (b,c) and As (d,e) diffractograms reveals systematic differences, both in the measured and calculated patterns. Arrows with numbers mark characteristic features. As eye-catching signatures, we find two arcs (1) in dark gray, running from left to right for Ga and from top to bottom for As (1`). These arcs (1) and (1`) appear in the calculations (c,e) in the same orientations as in the experiment, i.e., rotated by 90° with respect to each other. This is a clear fingerprint of the different orientation of the $Ga^{3+}$ and $As^{3-}$ coordination tetrahedra; see the sketch in Fig. 4(a). A second fingerprint are bright horseshoe-shaped features (2,2`) appearing in the four diagonal directions, which touch the arcs close to the rim of the field-of-view. The horseshoes are oriented close to horizontal for Ga (b,c) and close to vertical for As (d,e). Another characteristic are the faint dark crosses (3,3`) in the center of the diffractograms (b)–(e), whose orientations are rotated by 90°. All these details support that the images show a two-fold rotation symmetry, which is rotated by 90° between (b,c) and (d,e) due to the different orientation of the $Ga^{3+}$ and $As^{3-}$ coordination tetrahedra, marked in (a).

Figure 4 (g) shows the corresponding analysis for Si 2$p$, where the left and right halves show the measured and calculated patterns, respectively. This hXPD pattern shows fourfold symmetry, being indicative of the inversion-symmetric diamond-type lattice sketched in (f). All details in the measured and calculated diffractograms for GaAs and Si agree perfectly well, except for a slightly larger blur of the experimental patterns. This proves that for Si, Ga and As (Z=14, 31 and 33) the Bloch-wave approach with forward-scattering approx. constitutes an excellent model description with high predictive power.

*3.2 Observation of the structural phase transition in strontium titanate*

Strontium Titanate ($SrTiO_3$), a complex and most widely studied oxide with perovskite structure is a prototypical material for our understanding of structural phase transitions. At $T_c \approx 105$ K a transition from a low-temperature tetragonal phase (point group 4/mmm) to a high-temperature cubic phase (point group m3m) occurs. In this transition, the Ti-$O_6$ octahedra that make up its structure rotate in an alternating pattern, which is termed an 'antiferrodistortive' transition. $SrTiO_3$ was the first system for which experiments showed a soft mode, non-classical exponents, two time scales and two length scales for the fluctuations above $T_c$.[85]

This structural change in $SrTiO_3$ has become a model for similar transitions in a large number of compounds, occurring via mode softening. Upon substitution of different elements into $SrTiO_3$ the transition temperature shifts over a wide range, downwards (toward 0 K) or upwards (to above room temperature), the trends being quite complex. The key concept of understanding is the strain introduced by the dopant ions, more specifically the 'charge strain' or 'ionic valence mismatch' in the tetragonal

phase.[86]

Figs. 5(a-c) show the Sr $3d_{5/2}$, Ti $3p_{3/2}$ and O $1s$ diffractograms measured for $E_{final}$= 5.996 keV in the cubic phase (300 K). The different binding energies are accounted for by setting the appropriate photon energies according to eq. (1): 6.120 keV, 6.019 keV and 6.529 keV, for Sr $3d$, Ti $3p$ and O $1s$, respectively. Experiment (right halves) and calculation (left halves) show excellent agreement. The patterns are rich in details and their differences again reflect the different positions of the emitter atoms in the lattice, sketched in Fig. 5(d). The prominent crossing features, best visible and marked by arrows in the oxygen pattern (c), correspond to the <220>-type Kikuchi bands, confined by dark lines with distance $2G_{110}$ = 3.184 Å$^{-1}$.

The phase transition is related to the positions of the TiO$_6$ octahedra, which in the cubic phase are perfectly aligned with the Sr and Ti lattice [scheme Fig. 5(e)]. In the tetragonal phase, these octahedra are slightly rotated about one of the cubic main axes. Experimental values for the rotation angle range between 1.1° and 1.6° (cf. Ref. 87 and references therein). In adjacent unit cells they rotate in opposite directions [Fig. 5(f): view along (001), rotation angles exaggerated]. The lattice parameters change only slightly, i.e. the symmetry reduction from cubic to tetragonal is created by a <u>different, rotated, arrangement</u> of the O unit cell motif atoms in an otherwise similar (practically cubic) Bravais point lattice. Because the O atoms are also the weakest scatterers, the effect of their rotation would show up only relatively weakly in a non-element-resolved electron diffraction experiment (see the structure of SrTiO$_3$ by CBED in Ref. 87). If, however, we can analyze the specific emission from O relative to the surrounding Ti and Sr atoms as is possible with hXPD, the sensitivity to structural changes should be increased. This also means that the Ti and Sr emission in hXPD should show less effects, because their individual sublattices are less affected by the O rotation.

At first sight, the diffractograms at 30 K (not shown) look similar to those at room temperature. However, the low-temperature over room temperature (LT/RT) ratio images Figs. 5(g-i) and the detail in (p) uncover pronounced differences. Dark and bright small details are present in the Sr $3d$ and Ti $3p$ ratio patterns (g,h), uncovering a non-trivial appearance of the phase transition in the hXPD patterns. For O $1s$ the ratio pattern (i) exhibits even differences in the Kikuchi bands, breaking the fourfold cubic symmetry.

The microscopic effect of rotation of the oxygen emitter atoms in the $O_6$ octahedra is clearly visible in the macroscopic O 1*s* hXPD pattern. As expected, the O1s diffractogram shows the strongest differences.

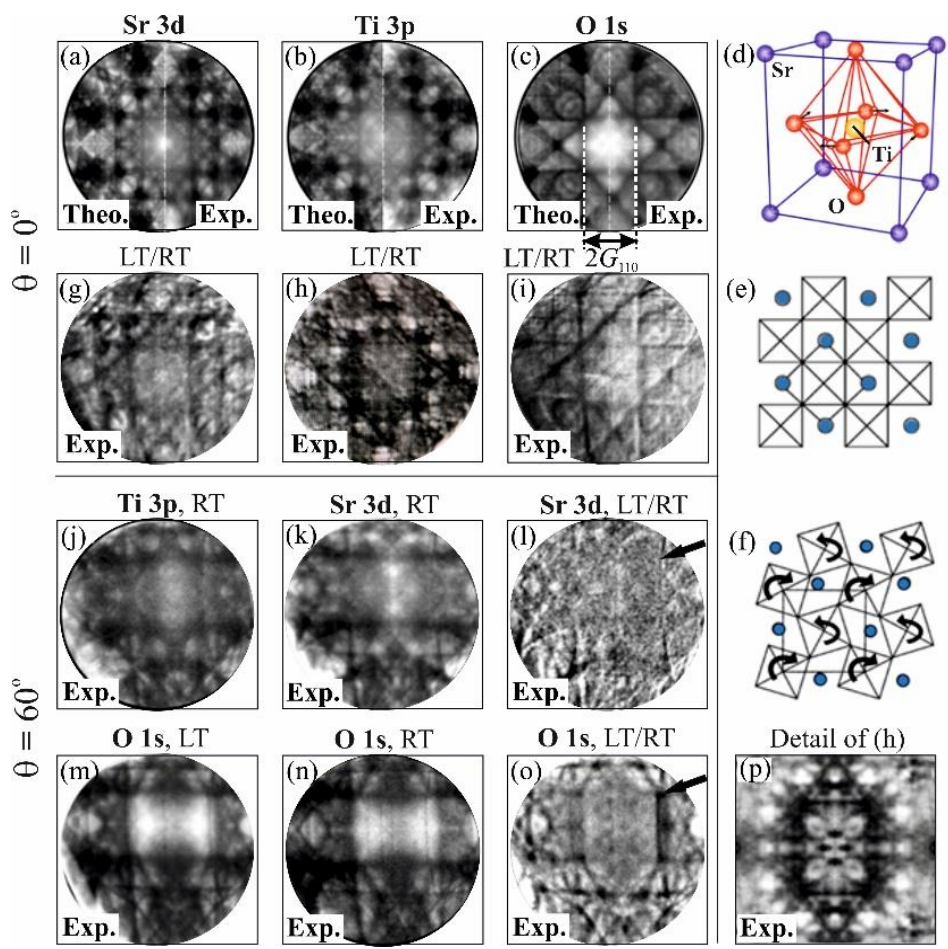

**Fig. 5.** Analysis of hXPD patterns for SrTiO$_3$ showing the cubic to tetragonal phase transition. (a-c) hXPD patterns showing the excellent agreement between measured (right halves of images) and calculated diffractograms (left halves) for Sr 3$d_{5/2}$, Ti 3$p_{3/2}$ and O 1*s*. Identical kinetic energies $E_{final}$=5.996 keV have been ensured by setting the appropriate photon energies. (d-f) Lattice model of SrTiO$_3$ for the cubic (d,e) and tetragonal phase (f). (e,f) Looking at (001) face; rotation angle of the Ti-O$_6$ octahedra is exaggerated. (g-i) Low-temperature / room-temperature (LT/RT: 30K / 300K) ratio images corresponding to the experimental patterns in (a-c). (j-o) Off-normal diffractograms at polar angle θ=60° (referring to center of pattern), recorded in zero-field mode. (j,k) hXPD patterns for Ti 3*p* and Sr 3*d*, (l) LT/RT ratio image for Sr 3*d*. (m,n) LT and RT hXPD patterns at the O1*s* line, the differences are visible without data processing. (o) LT/RT ratio image uncovering a rich pattern of differences upon crossing the phase transition. (p) Detail of (h) after symmetrization.

For such complex structural changes, the near-normal emission region might not be ideal for viewing the most significant changes. Hence, we took further patterns in strongly off-normal direction centered at θ=60° [Figs. 5(j-o)], using the zero-extractor-field mode. The quality and richness in details of the 60° patterns is comparable to the 0° patterns, proving that this zero-field mode does not sacrifice image quality. Only the size of the observed k-field is slightly smaller.

Figs. 5(j,k,l) show the Ti 3$p$ and Sr 3$d$ patterns and the ratio (LT/RT) for Sr. Figs. 5(m,n) show the LT and RT patterns for O 1$s$ and here differences are discernible without data treatment (e.g. vertical lines in the bright rectangle). The LT/RT ratio pattern (o) exhibits pronounced features, again involving the borders of the Kikuchi bands (dark lines) and details in the central zone. Indeed, the largest differences are captured in off-normal emission direction and for the oxygen emitters.

X-ray diffraction at the Ti-Kα line[88] revealed that between 30 K and 300 K the lattice parameters show a variation in the range of $1.8 \times 10^{-3}$ and that the evolution of the lattice parameters with temperature in the near surface region is changed with respect to the bulk. hXPD would allow varying the depth of the probe volume by setting high and low kinetic energy, which is beyond the scope of the present survey study. The measurements of Fig. 5 primarily addressed the question, how the phase transition shows up in the patterns and whether off-normal emission can enhance the visibility of LT/RT differences. The element-specificity of hXPD enables a more detailed view into the high-energy diffraction mechanisms. Likewise, a lattice-site analysis of substitutional atoms could shine light into their role of shifting $T_c$ as studied in Ref. 86.

A detailed theoretical analysis of the tetragonal phase is in progress and will be subject of a forthcoming publication. Experimental studies with much higher spatial resolution will become feasible in the near future, when a focusing X-ray capillary will be implemented at beamline P22. This will allow us to resolve the domain structure of the tetragonal-distorted phase. A Bloch-wave calculation[6] for Si in the cubic phase and with uniaxial tetragonal compression revealed strong changes in the fine structure of the central 'zone' (crossing region of the horizontal and vertical Kikuchi bands). Our SrTiO$_3$ results also show characteristic changes in the central zone, see Figs. 5(g-i) and (p), showing the enlarged central detail of (h). Likewise, the off-normal zones in (l,o), marked by arrows, show fine-structure patterns arising upon crossing of the cubic to tetragonal transition temperature. This first pilot experiment, still lacking systematics (e.g. varying the photon energy as in Ref. 56) gives a first glance on the potential of hXPD for studies of phase transitions.

*3.3 Analysis of dopant and impurity sites*

Here we address an issue of practical importance for the analysis of XPD patterns: the emitter-site specificity, which can be exploited for the localization of dopant atoms or impurities in host materials. Pioneering work, extracting detailed information on dopant sites from XPD / photoelectron holography has been performed by Kato et al.[89, 90] and Tsutsui et al.[91] In the former papers, atomic stereophotography was used to study the dopant sites of boron in diamond, with emphasis on the surface specificity, revealing a difference between the (111) and (001) surface. In the latter work an implementation of a high energy resolution electron analyzer with spectro-photoelectron holography for 3D imaging of local dopant atomic structures of As-doped Si and first-principles simulations are presented. The challenge was to combine high energy resolution with high sensitivity simultaneously.

The element specificity of XPD can be exploited to determine the position of dopant atoms within the bulk of a host lattice, with a selected core-level signal serving as spectroscopic fingerprint. Information on dopant sites is of particular importance for semiconductor design because simultaneous occupation of different sites can take place, some of them counteracting the desired hole or electron doping effect.[92] There is a wealth of previous studies concerning dopant sites, mostly using ion channeling,[93] Rutherford backscattering,[94, 95] near-edge x-ray absorption fine structure,[96] infrared photoresponse,[95] or diffraction of beta-rays from implanted radioactive nuclei,[97, 98] indicating the importance of this task. XPD adds complementary information on the specific site within the translational-symmetric lattice that can only indirectly be deduced from the above-mentioned methods.

As discussed in 3.1, hXPD offers the possibility to distinguish the Ga and As sites via an hXPD analysis exploiting the non-centrosymmetric structure of the zinc-blende lattice of GaAs. Here we apply this site sensitivity for a dopant analysis in (In,Ga,Mn)As films with 3% In plus 2.5% or 5.6% Mn[17]. ToF-MM yielded information on the dopant sites and revealed a shift of the chemical potential with increasing Mn doping and a highly-dispersing band, crossing the Fermi level for high Mn concentration.

Figure 6 presents the diffraction patterns of the four constituents measured at the final-state energy of 3.31 keV. Dotted lines and numbers mark characteristic features.

Inspection of the Mn 3*d* hXPD pattern (c) reveals that the arc (1) runs from left to right, and the horseshoe with a center cross is visible on the top and bottom. This is the signature of the Ga substitutional site.

The second row of Fig. 6 shows calculated Kikuchi diffractograms for Mn atoms on Ga (e) and As sites (f), which are identical to the Ga and As patterns at the same final-state energy. The signatures are in accordance with the experimental patterns (a), (b). They show the same orientation of arcs (1) for Ga and (1`) for As, of the small horseshoes along the diagonal (2,2`), and of the larger horseshoes with a cross (3,3`). Incorporation of Mn on the substitutional Ga sites is also evident from the comparison of (c) and (e). In addition, the Kikuchi fingerprints of Mn atoms on tetrahedral and hexagonal interstitial sites were calculated [see Figs. 6(g,h), respectively]. It is clearly seen that the hexagonal site is characterized by dark diagonal Kikuchi bands of {001}-type that cross the horizontal {011}-band.

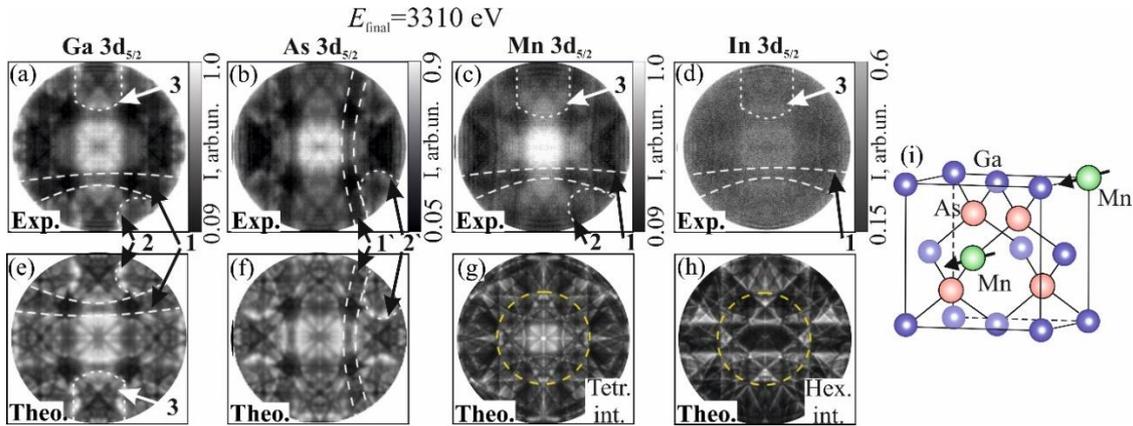

**Fig. 6.** (a)–(d) hXPD patterns of photoelectrons from the $3d_{5/2}$ core levels of Ga (a), As (b), Mn (c) and In (d) of an epitaxial (In,Ga,Mn)As thin film with nominal doping of 3% In and 5.6% Mn. The kinetic energy was set to 3.31 keV by tuning the photon energies to hν = 3317, 3340, 3355, and 3342 eV for (a, b, c and d). (e)–(h) Kikuchi diffraction calculation for Mn atoms on Ga (e) and As sites (f) as well as for Mn atoms on tetrahedral (g) and hexagonal interstitial sites (h), all for $E_{kin}$ = 3.31 keV. Dotted lines and numbers mark characteristic features; the dashed circles in (g), (h) denote the field-of-view of (a)–(f); (i) GaAs crystal structure with Mn atoms in Ga substitutional sites. (from Ref. 17).

The crossing region in the center appears very dark. This is not compatible with the bright center region of the measured diffractogram (c). The tetrahedral interstitial

(g) shows a bright center region but the rest of the pattern does not resemble the experimental pattern (c). Comparison of experimental and calculated patterns confirms that Mn predominantly occupies substitutional Ga sites, even in the hyperdoping regime.

Thanks to their predictive power, hXPD calculations of different configurations in comparison with experiment constitute a new tool for a detailed analysis of the site-distribution of dopants and impurities. This approach has been used for the case of Te atoms in a Si crystal.[16, 99] In order to determine the prevailing sites, systematic calculations were carried out for many different configurations. Figure 7 shows a few examples, demonstrating how this method works in practice. As an example, a measured Te 3$d$ diffractogram for a sample with 5.6 at% Te is shown in comparison with various calculations (all at $E_{final}$ = 5.69 keV). The experimental Kikuchi pattern of the dopant (a) shows excellent agreement with the calculated pattern (b) for individual Te-monomers in substitutional sites (statistical sum of sites A and B). The calculations for the tetrahedral (c) and hexagonal (d) interstitial sites predict markedly different Kikuchi patterns. Hence, these sites can be ruled out as explanation for this experimental pattern. Close inspection of the chemical shifts of the 3$d$ signal uncovers several coexisting Te sites; for further details, see Refs. 15 and 99.

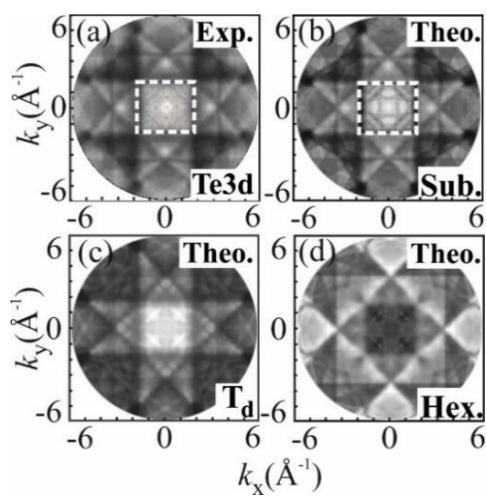

**Fig. 7.** Measured Te 3$d$ hXPD diffractogram of Si(001) hyperdoped with 5.6% Te (a) in comparison with Kikuchi patterns calculated for substitutional sites (b), Te statistically-distributed on tetrahedral (c) and hexagonal interstitial sites (d). The central zones in (a) and (b) are shown with different contrast in order to avoid saturation. The final-state energy is 5.69 keV (from Ref. 16).

In addition to the structure visible in the diffractograms, the absolute contrast in the patterns bears information, cf. scale bars in Figs. 6 (a-d). Experimentally, there are several intrinsic and extrinsic sources of diffuse background. Intrinsic sources are thermal diffuse scattering, disorder in the lattice and the fact that core-level signals are superimposed by a contribution from the secondary-electron cascade. Extrinsic sources are the noise background of the MCPs, a possible non-linearity in MCP response (depending on count rate and how evenly the MCP is illuminated) and scattering processes of electrons in the microscope column (which can be avoided by appropriate aperture settings). All these contributions would reduce the absolute value of the image contrast in hXPD patterns. On the theoretical side, calculations would require inclusion of inelastic effects and Debye-Waller factors, which is beyond the scope of the current paper.

In this case, the calculations have been performed *a posteriori*, hence there was no pre-selection of 'best' conditions in parameter space (final-state energy and angular range). Ideally, the calculations should be done *a priori*, so that the expected signature is most pronounced. This survey calculations helped to find such optima for the case of the Si host lattice. For other lattices, new calculations are required. The paper (Ref. 16) was intended as theoretical proof-of-concept and first implementation of hard x-ray photoelectron Kikuchi-diffraction for emitter-site specific structural analysis. The main result of many calculations, such as interstitial sites (tetrahedral and hexagonal), fully arbitrary sites, multimers and clusters show strongy different hXPD signatures.

## 4.     XPD in valence-band momentum patterns

Photoelectron diffraction in valence-band (VB) photoemission is more complex than for core levels. The roles of band dispersion, initial electron momentum and photon momentum in VB-XPD is not *a-priori* clear. Early studies at high energy and high temperature [i.e. strong quasi-elastic scattering, eqs. (3,4)] revealed matrix-element weighted densities of states (MEWDOS), modulated by XPD effects.[100, 101, 102] The results could be interpreted analogously to core-level XPD after integrating over an energy range of a few eV [103] or by the assumption of a localization of the band electrons[100] as explained in orthogonalized-plane-wave approximation.[104] In ToF-MM data, valence-band XPD appears in terms of strong local intensity enhancements by up to a factor of 5 in small intervals of ($E_B$,**k**) parameter space ($\Delta k$-regions down to 0.03 Å$^{-1}$; $\Delta E_B$ down to 200 meV), overlaid on the valence-band patterns. The different classes of Laue- and Kikuchi-type diffraction in VB-XPD has been discussed in detail[76]; here we recall the essentials and give a selection of typical examples.

Considering a direct interband transition in 3D k-space, all final states lie on an energy isosphere, the diameter of which is determined by energy conservation [eq. (1)]. This final-state sphere cuts through the periodic band patterns in k-space. The intensity distribution on the intersection contour is directly observed in a momentum microscope.[30,105] Most important are the different roles of the two relevant energies, $E_{final}$ and $E_B$. The former defines the wavelength of the photoelectron, which governs diffraction dynamics. The latter defines the relevant isosurface of the spectral function $\rho(E_B,\boldsymbol{k})$, which is strongly energy dependent, giving rise to a similar energy dependence of the intensity modulation on the eV-scale. In core-level XPD $E_B$ is fixed and there is no $E$-vs-$\boldsymbol{k}$ dispersion, hence only one energy- and one momentum-condition exist.

Involving reciprocal lattice vectors, VB photoemission is understood as an umklapp process. Momentum conservation demands that the photon momentum enters into the momentum balance. Hence the description of a VB-XPD process in k-space is rather complex. It can intuitively be understood as umklapp process on the final-state energy isosphere. The transfer of photon momentum causes a shift of the centre of this isosphere; for details, see ref. 76.

VB-hXPD causes an intensity modulation of the valence-band pattern. Babenkov et al.[106] observed, that the modulation imprinted on the valence-band pattern strongly resembles core-level hXPD patterns recorded at the same $E_{final}$ (i.e. photoelectron kinetic energy inside of the material). This opens a path to a multiplicative correction, eliminating the imprinted modulation.

Fig. 8 shows three examples of as-measured VB momentum patterns for TiTe$_2$ (a), GaAs (f) and graphite (k). All three k-images are dominated by imprinted hXPD patterns. This becomes evident when recording core-level hXPD patterns at the same final-state energy (in order to ensure identical photoelectron wavelengths and exclude dynamical effects). The core-level diffractograms of Ti 4$d_{5/2}$ (b), As 3$d_{5/2}$ (g) and C 1$s$ (l) exhibit the same characteristic patterns as imprinted on the valence maps (a,f,k). This modulation can be largely eliminated via pixel-by-pixel division of the as-measured VB distribution by the core-level hXPD pattern. Indeed, this numerical correction leads to the true VB momentum distributions as visible in (d, h and m). Owing to the large k-field, these images show the periodic patterns of many BZs. Further data treatment like background subtraction, averaging and symmetrisation allow extracting a single BZ pattern in high quality, e.g. (i). Note that (d, h, i and m) show only one out of many energy slices; the band dispersions for the three examples are shown in (e, j and n). For details on the TiTe$_2$ and GaAs results, see Refs. 106 and 17, respectively.

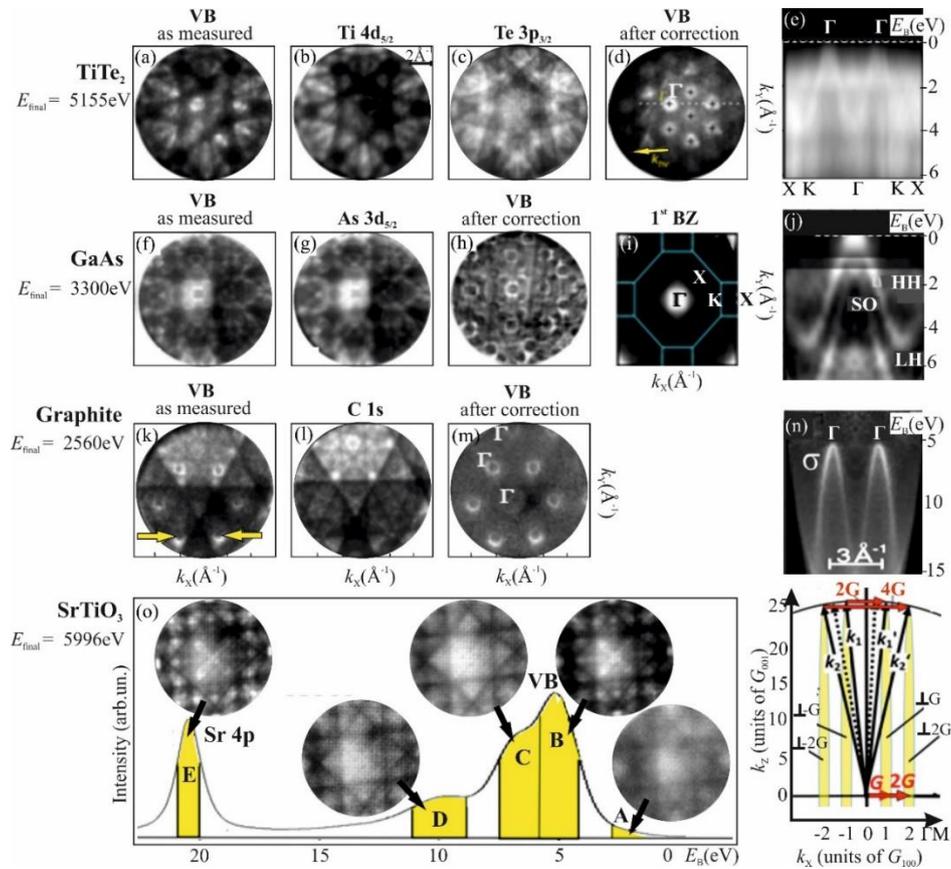

**Fig. 8.** Examples for valence-band XPD for TiTe$_2$ (a-e), GaAs (f-j), graphite (k-n) and SrTiO$_3$ (o). (a,f,k) As-measured VB momentum patterns; final-state energies given in the panels. (b,c,g,l) Core-level hXPD patterns recorded at identical $E_{final}$; core-levels denoted in the panels. (d,h,m) Corrected valence-band patterns, after pixel-by-pixel division by the core-level diffractograms; (i) single Brillouin zone after averaging and background subtraction. (e,j,n) Corresponding band dispersions after XPD elimination. (o) Five hXPD patterns for SrTiO$_3$ in the valence range and shallow Sr 4$p$ core level, revealing different diffraction signatures in different regions A-B (marked in yellow) of the valence band. (p) Schematic k-space model of Kikuchi diffraction. [(a-e) from Ref. 106, (f-j) from Ref. 17]

The graphite pattern (k) shows clearly, how the dark horizontal Kikuchi band 'cuts' through two of the σ-band circles (marked by the arrows). This proves that the band features themselves experience Kikuchi-type diffraction. In addition, the as-measured VB patterns show a strong background caused by quasi-elastic diffuse scattering. As discussed in Sec. 2.3 this background shows a Kikuchi pattern as well, with the wave-field of the scattered wave centered at the scatterer. Pixel-by-pixel division of Fig. 8(k) by the C 1s pattern (l) completely removes the superimposed Kikuchi pattern, yielding the true VB momentum distribution Fig. 8(m), i.e. the circular sections through the graphite σ-band with its hole-like dispersion shown in (n).

This kind of correction has been successfully applied to a number of systems studied with hard X-rays and has become a standard data-treatment routine. In fact, it is necessary to avoid misinterpretation of HARPES results. Next to the examples shown in Fig. 8,[17, 106] ToF-MM-HARPES captured a Néel-vector-induced change of valence states in the collinear antiferromagnet $Mn_2Au$,[107] the temperature dependence of the Fermi surface and band structure of the Kondo-material $YbRh_2Si_2$ [108] and the bulk band structure of the Heusler compounds $Co_2MnGa$ and $Co_2MnSi$[109].

*The background in a VB pattern thus contains valuable information on the scattering site.* Close inspection of Figs. 8(a-c) reveals that the Kikuchi pattern imprinted in the VB (a) strongly resembles the Ti 4d diffractogram (b), whereas the hXPD signature of Te 3p looks markedly different (c). We can infer from this difference that the wavefunction of the valence state is mainly located *at the Ti atoms*.

We have looked at this phenomenon for $SrTiO_3$ in a fast survey study shown in Fig. 8(o). At the final-state energy of almost 6 keV and room temperature, the thermal-diffuse scattering is very strong. Thus the valence bands are invisible in the as-measured patterns. At such conditions it is easy to capture the hXPD signatures of various regions A-D, which look quite different. The yellow areas denote the regions integrated for the hXPD diffractograms. The pattern taken in the high-energy wing in the band gap close to the Fermi energy (region A) shows only a weak contrast (right hXPD pattern). Pronounced patterns are observed in the three VB regions B, C and D and in the shallow core level Sr 4p at $E_B \approx$ 20.5eV (E, left pattern). The highest VB-peak (B) clearly shows the signature of strontium as visible in the Sr 4p pattern (E). However, shoulder C and peak

D carry the signature of oxygen [cf. Fig. 5(c)]. For a detailed discussion of the valence-band spectrum of SrTiO$_3$, see Ref. 110. We conclude that hXPD enables to probe, where a valence-band wavefunction is localized. Fig. 8(p) shows a k-space scheme of Kikuchi diffraction, resulting from reflection at sets of lattice planes perpendicular to the reciprocal lattice vector **G**.

## 5. Conclusions and future prospects

This article reviews hard-X-ray photoelectron diffraction (hXPD) for geometric structure analysis, using a high-energy variant of time-of-flight momentum microscope (ToF-MM). This instrument is capable of recording high kinetic energies up to >7 keV and full-field momentum images with up to 20 Å$^{-1}$ diameter. Various examples are discussed, emphasizing the complementary nature of hXPD and holographic XPD at lower energies. Parallel ($k_x,k_y$)-imaging and ToF energy recording effectively counteract the rapidly dropping cross sections and rising phonon-scattering probability with increasing photon energy in the hard X-ray range.

hXPD is characterized by a large IMFP and hence a large probe volume, leading to rich patterns with filigree fine structure. Given atom numbers of typically several 10$^6$ in the effective probe volume, the diffraction process can be well described by Bragg reflection at lattice planes. The hXPD scenario can be interpreted as 'half a Kikuchi process', where the initial inelastic scattering step from an incident electron beam is missing. In turn, calculations using the Bloch-wave approach to XPD yield a stunning agreement with experiment.[56] This sensitivity on the long-range order is nicely complementary to the photoelectron holography approach at energies <1 keV, where the local surroundings of the emitter atom govern the diffraction pattern.[10, 11] Photoelectron holography can capture the local 3D atomic configuration and is principally applicable to non-periodic structures. The quality of reconstructed atomic images has greatly improved thanks to the development of a new extraction algorithm using the maximum entropy method / L1 regularization[11,41].

Despite its long-range sensitivity, local information can be retrieved from hXPD patterns as well. Thanks to the short photoelectron wavelength (e.g. 14.6 pm at 7 keV) the patterns change strongly, when the emitter atom is moved by a fraction of an Å within the unit cell, as validated in model calculations[16] For instance, calculations for various emitter sites can be used as fingerprints for dopant positions. At high energies, we can identify different dopant sites by their hXPD fingerprints, whereas in photoelectron holography the real-space atom pattern is extracted by a transformation routine.

In order to understand what we see in hXPD, we looked deeper into the nature of the photoemission final state. The observed photoelectron wave originates from a transition between an initial state (e.g. a selected core level) and a spherical-wave, which is diffracted at the lattice. This 'multiple-scattering' photoemission final state (time-reversed LEED-state) was first discussed by Pendry[61]. The diffraction dynamics depends sensitively on the position of the emitter atom in the lattice, being generally different for the various constituents in compounds. Measured diffractograms of outgoing photoelectron waves for Nb $4p_{3/2}$ and Se $3p_{3/2}$ in $NbSe_2$ and for Sr, Ti and O core levels in $SrTiO_3$ are markedly different. This aspect goes beyond the classical Kikuchi effect in electron microscopy, which is not observed element selectively. In order to exclude dynamical effects, such images are recorded at identical kinetic energies by choosing the appropriate photon energies.

Quasi-elastic *thermal diffuse scattering* events kick electrons out of the final-state wave field. The scattered electrons are by no means 'diffuse' but constitute a new wave field centered at the scattering site. Such scattering events gain importance with increasing kinetic energy and increasing temperature, as governed by the Debye-Waller factor. The diffuse-scattering scenario resembles the classical Kikuchi process, with the only difference that the initial wave is the Pendry final state, instead of the plane wave of the incoming beam in a TEM or SEM.

We conclude by recalling some of the special features of hXPD: Unlike conventional diffraction, XPD circumvents Friedel's law (stating that the squared Fourier amplitude is centrosymmetric). For epitaxial GaAs thin films (zinc-blende structure) the missing inversion centre is clearly visible in the Ga and As diffractograms, proving that the zinc-blende structure is preserved in the film. Thanks to the small photoelectron wavelength, structural phase transitions lead to marked changes in the hXPD pattern as shown for the cubic-to-tetragonal transition in $SrTiO_3$. Calculations using the Bloch-wave approach can predict the fingerprint-like signatures of any desired emitter site in a host material, as shown for Mn in GaAs and Te in Si. Combining core-level hXPD with hard-X-ray ARPES (HARPES), it is possible to eliminate the strong diffraction signature imprinted in bulk band maps taken with hard X-rays. Identical settings (kinetic energy, *k*-scale) enable a multiplicative evaluation / correction procedure (ratio images, removal of imprinted

Kikuchi modulations) of the patterns. The imprinted pattern can yield information on the localization of the corresponding valence-band wavefunction.

As an **outlook,** we address upcoming progress on the instrumental side opening the path to new exciting scientific opportunities. The zero-field mode of the front lens as discussed in Sec. 2.1 and used for the measurements Figs. 5(j-o) enables non-planar sample geometries (e.g. cleaved microcrystals) and complex sample mounts. Such 3D-structured samples are prohibitive for conventional PEEM, LEEM or low-energy MM applications due to the required extractor field. Electrical contacts or actuators on top of the sample mount will allow *measurements in-operando or under mechanical strain*. Going to even higher kinetic energies will allow measurements for liquid films under pressure, where the separation from the vacuum side is done by a thin electron-transparent foil in the beam path. A practical application of hXPD with high potential is dopant-site analysis e.g. in semiconductors, exemplified for Mn in GaAs and Te in Si (further results have been obtained for Ga in Ge). Recent experiments revealed large circular dichroism asymmetries (up to >80%) with surprisingly strong structure in hXPD patterns at hν= 6 keV for low-Z and high-Z materials (Si and W, respectively). We expect a similarly strong hXPD spin-polarization texture for high-Z materials, excited with circ. polarized X-rays.

ToF-MM exhibits its main advantage in cases, where weak signals need to be detected and discriminated from a larger background. A recent example is time-resolved (tr) XPD for the observation of ultrafast processes involving the lattice. Such experiments are becoming possible due to the availability of fs-pulsed X-ray beams from FELs with high pulse rate. Rapid switching between XPD and (H)ARPES is possible just by changing the photon energy, tuning the core-level XPS peak to the same kinetic energy as the valence-band signal. First experiments using higher-harmonic-generation (HHG) sources have impressively demonstrated the potential of tr-ToF-MM for the study of ultrafast processes.[111,112,113,114,115,116,117] The 3D ($E_{kin}$,$k_x$,$k_y$) recording scheme is advantageous in particular, when it is not *a priori* known, where in k-space the interesting physics is going to happen. A key component of such experiments is the detector, since non-optimized detector configurations (like the missing high-pass filter in the MM-experiment discussed in Ref. 118) can strongly reduce the overall performance.

Combining pump pulses in a wide range from the visible to the THz spectral range with X-ray probe pulses enables to track the time-evolution of phenomena involving electronic system and lattice (like the coupling with coherent phonon excitation) after fs excitation. In parallel to the progress using HHG sources, tr-ToF-MM was established at the FEL FLASH at DESY, Hamburg.[45,46,47,48] Photon energies up to 750 eV enable tr-XPS and tr-XPD, probing structural changes on femtosecond timescales. Simultaneous core-level and conduction-band studies enabled to capture a Mott transition through the 'eyes of a core level',[47] using the transient linewidth broadening for detecting the electron temperature[46] or ultrafast lattice vibrations in $Bi_2Se_3$ studied via excitation of coherent phonons (A1g symmetry) by fs linearly polarized optical pulses.[49] Two further ToF-MM projects that will use fs X-rays up to several keV are just being launched at the European XFEL[119] and at LCLS-II in Stanford.[120]

**Acknowledgments**

The results shown in this paper have been obtained in cooperation of the group at Mainz University with several partners. Sincere thanks go to Katerina Medjanik, Dmitri Vasilyev, Sergey Babenkov, Sergey Chernov, and Hans-Joachim Elmers (Mainz University), to Kai Rossnagel and coworkers (Kiel University) for providing the $NbSe_2$ sample, and Ralph Claessen, Judith Gabel and coworkers (Würzburg University) for providing the $SrTiO_3$ sample. For excellent support, we thank the staff of beamline P22 at PETRA-III, in particular Christoph Schlueter, Thiago Peixoto and Andrii Hloskovsky and the HEXTOF team at FLASH, in particular Dmytro Kutnyakhov, Markus Scholz, Michael Heber and Nils Wind. Special thanks go to Moritz Hoesch for many discussions on hXPD and photoemission in general and a critical reading of the manuscript. The ToF-MM experiments were funded by BMBF (05K16UM1, 05K16UMC, 05K19UM1 and 05K19UM2) and Deutsche Forschungsgemeinschaft DFG (German Research Foundation) through TRR 173–268565370 Spin+X (project A02). A.W. was supported by the Polish National Science Centre (NCN) grant number 2020/37/B/ST5/03669.

**Author contributions**
O. F. and A.W. performed the Bloch-wave calculations. All three authors have written the paper and made the evaluation of the new data in Figs. 2, 5 and 8.

**ORCID iDs**
O Fedchenko https://orcid.org/0000-0002-6159-7934
A Winkelmann https://orcid.org/0000-0002-6534-693X
G Schönhense https://orcid.org/0000-0002-8921-2901

*E-mail: schoenhe@uni-mainz.de